\documentclass{article}

\usepackage[final]{neurips_2024}
\usepackage{amsmath,graphicx}
\usepackage{subfigure}
\usepackage{siunitx}
\usepackage{colortbl}
\usepackage[table,xcdraw]{xcolor}

\usepackage[utf8]{inputenc} 
\usepackage[T1]{fontenc}    
\usepackage{hyperref}       
\usepackage{url}            
\usepackage{booktabs}       
\usepackage{amsfonts}       
\usepackage{nicefrac}       
\usepackage{microtype}      
\usepackage{xcolor}         
\usepackage[numbers]{natbib}

\definecolor{nmcolor}{RGB}{194,81,48}

\newcommand{\model}{\textsc{DART}}
\newcommand{\encoder}{\textsc{ML-VAE}}
\newcommand{\baselineA}{\textsc{MLVAE-Taco}}
\newcommand{\baselineB}{\textsc{DART}$_{scratch}$}
\newcommand{\baselineC}{\textsc{DART}$_{w/o~VQ}$}
\newcommand{\baselineD}{\textsc{Multispk-FS2}}
\newcommand{\baselineE}{\textsc{GST-FS2}}
\newcommand{\baselineF}{\textsc{GST-GE2E-FS2}}

\title{DART: Disentanglement of Accent and Speaker Representation in Multispeaker Text-to-Speech}

\author{%
  Jan Melechovsky$^*$, Ambuj Mehrish$^*$, Berrak Sisman$^\dagger$, Dorien Herremans$^*$ \\
  $^*$Audio, Music, and AI Lab, Singapore University of Technology and Design, Singapore\\
  $^\dagger$Speech \& Machine Learning Lab, The University of Texas at Dallas, USA\\
  \texttt{jan\_melechovsky@mymail.sutd.edu.sg} \\
}

\begin{document}

\maketitle

\begin{abstract}

Recent advancements in Text-to-Speech (TTS) systems have enabled the generation of natural and expressive speech from textual input. Accented TTS aims to enhance user experience by making the synthesized speech more relatable to minority group listeners, and useful across various applications and context.
Speech synthesis can further be made more flexible by allowing users to choose any combination of speaker identity and accent, resulting in a wide range of personalized speech outputs.
Current models struggle to disentangle speaker and accent representation, making it difficult to accurately imitate different accents while maintaining the same speaker characteristics. We propose a novel approach to disentangle speaker and accent representations using multi-level variational autoencoders (ML-VAE) and vector quantization (VQ) to improve flexibility and enhance personalization in speech synthesis.
Our proposed method addresses the challenge of effectively separating speaker and accent characteristics, enabling more fine-grained control over the synthesized speech.
Code and speech samples are publicly available\footnote{\url{https://amaai-lab.github.io/DART/}}.
\end{abstract}

\section{Introduction}
\label{sec:intro}

In recent years, Text-to-Speech (TTS) technology has advanced significantly, allowing high audio quality synthesis in multiple voices for applications such as voice assistants, audiobooks, and entertainment \cite{mehrish2023review}. Despite their advancements, a significant challenge remains: effectively disentangling speaker identity and accent representations to achieve precise and personalized speech synthesis.
With globalization, accents in speech technology are vital for effective communication, since a listener's ability to understand a speaker is determined by both the speaker's accent and the listener's familiarity with that particular accent \cite{wells1982accents}.
However, expecting everyone to learn a single standard accent is impractical. Instead, we should focus on developing technologies that can generate accents according to the user's needs.
Accents involve phonetic and prosodic variations influenced by factors like mother tongue or region \cite{wells1982accents, lippi2012english}. Since accent forms a part of one's idiolect, it may often overlap with speaker identity \cite{wells1982accents}, which makes the disentanglement a challenge. Successfully disentangling the two elements would allow for personalized speech synthesis, improving user experiences for minorities by aligning the system's accent with their own to promote intelligibility, thus enhancing interaction with TTS voice assistants and audiobook narrators.


The introduction of deep learning to TTS pushed the research forward with models like WaveNet \cite{vanwavenet}, Tacotron \cite{wang2017tacotron,shen2018natural}, and Fastspeech2 \cite{fastspeech2}. Multi-speaker TTS systems have advanced this field further, enabling speech synthesis in different voices and styles by training on diverse datasets with recordings from multiple speakers \cite{gibiansky2017deep,xue2022ecapa,kim2021conditional}. These systems can mimic accents \cite{liu2024controllable,melechovsky2022accented,zhou2023tts} and emotional expressions \cite{im2022emoq,lei2022msemotts}. Continued research in multi-speaker TTS is expected to enhance synthesized speech quality and versatility.
However, previous studies (for related works, please see \ref{app:relworks}) have not thoroughly explored disentangling accent and speaker representation, which could unlock the potential to further improve personalization level and help promote underrepresented foreign accents. In similar work, Melechovsky et. al. \cite{melechovsky2023learning} proposed ML-VAE along with Tacotron2 to disentangle speakers and accents. However, the resulting accent similarity was not shown to be overwhelming and experiments were done on native accents of English only. Inspired by this work, we aim to expand on this idea to strive for even better disentanglement with focus on foreign accents.


In this paper, we propose \textbf{D}isentanglement of \textbf{A}ccent and Speaker \textbf{R}epresen\textbf{T}ation \model{}, which combines Multi-Level Variational Autoencoders (\encoder{}) \cite{bouchacourt2018multi} and Vector Quantization (VQ) \cite{van2017neural} to learn meaningful disentangled latent representations for speaker and accent.
The \encoder{} architecture forms the core of accent and speaker identity disentanglement. Through this variational framework, the model learns a latent space to represent the two, offering precise control during speech synthesis.
Furthermore, VQ discretizes the continuous latent variables obtained from the \encoder{}. This discretization maps the continuous latent space into a predefined codebook of discrete vectors, reducing the complexity of the latent space and promoting better separation of speaker and accent embeddings.
Through extensive experiments on diverse accented speech data \cite{zhao2018l2}, we evaluate the effectiveness of our proposed approach. Our major contributions are as follows: (1) We propose a novel architecture for disentangling speaker and accent representation using ML-VAE and VQ. (2) Through comprehensive experiments, we demonstrate the critical role of pre-training the TTS backbone on a multi-speaker English corpus for effective accent conversion and speaker/accent disentanglement.

\section{\model{}}
\label{sec:method}
\subsection{Backbone TTS Model}
\begin{figure*}
    \centering
    \includegraphics[width=0.8\columnwidth]{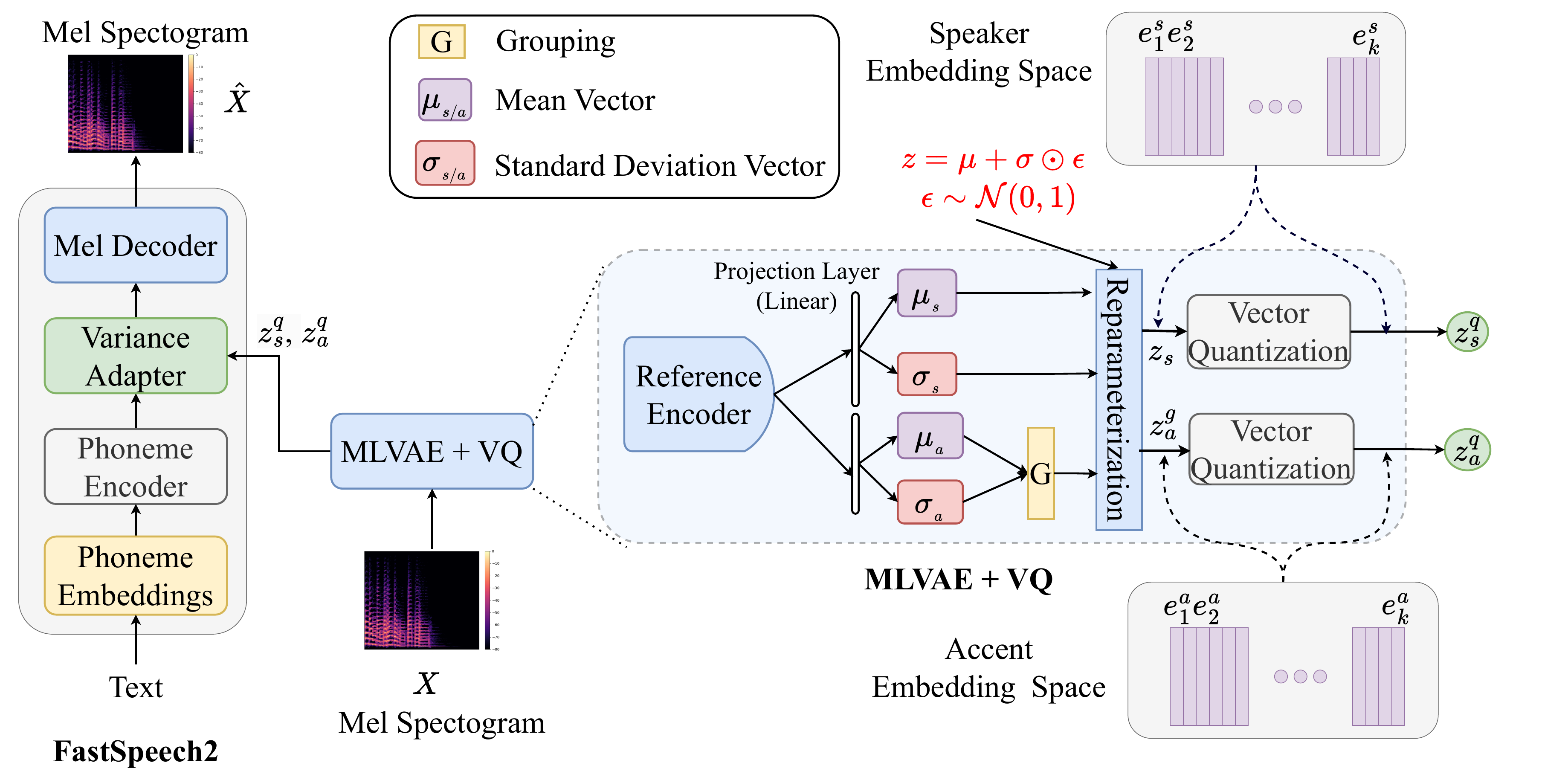}
    \caption{Architecture of \model{} including encoder, the \encoder{}, VQ, variance adapter, and decoder.}
    \label{fig:startts}
    \vspace{-5mm}
\end{figure*}

The first component of \model{} is the TTS backbone, which closely resembles Fastspeech2 architecture,
comprising of phoneme encoder, Variance Adapter, and Mel-Decoder, as depicted in Figure \ref{fig:startts}. To initialize the backbone, we perform pre-training on LibriTTS, an extensive multi-speaker dataset. The model is trained using the reconstruction loss between the predicted mel spectrogram $\hat{X}$ and the ground truth mel spectrogram $X$ is computed using Eq \ref{eq:lrecon}, where $||.||_{2}$  denotes $L_{2}$ norm.
\begin{equation}
\label{eq:lrecon}
    L_{recon} = ||\hat{X}-X||_{2}
\end{equation}

\subsection{ML-VAE Encoder}
\encoder{} \cite{bouchacourt2018multi} leverage the hierarchical structure of data to model the joint distribution of observed data and latent variables across multiple levels. This allows to encode dependencies among latent variables and disentangle different factors of variation in data generation. Additionally, it can utilize grouping information from real-world datasets, identifying shared variations and learning group-specific factors. This makes it ideal for datasets with natural grouping or clustering, such as categorical images or multi-sensor time series data. The architecture is based on variational inference techniques, which enable the learning of model parameters and efficient posterior inference. \encoder{} encodes disentangled representations of grouped observations characterized by different accents, examining their influence on underlying speech factors. To achieve latent representation separation, we utilize two variables: \(z_s\) for speaker-related variation and \(z_a^g\) for accent-related variation, where the superscript \(g\) denotes speaker grouping based on accent. These variables allow us to disentangle distinct factors of variation in speech data. During \encoder{} training, we optimize an objective function similar to previous work \cite{melechovsky2023learning}. The \encoder{} effectively captures salient variation factors in speech while disregarding irrelevant factors. Interested readers can refer to \cite{bouchacourt2018multi} for detail analysis of the \encoder{} architecture, implementation, and experimental setup. The KL loss \(\mathcal{L}_{kl}\) is computed by maximizing the group ELBO over mini-batches.

\subsection{Vector Quantization}
We extend the \encoder{} framework from \cite{melechovsky2023learning} by integrating VQ into a unified architecture, \model{}. Our design incorporates separate VQ modules for accent and speaker in the \encoder{} encoder (Figure \ref{fig:startts}, with codebook dimensions \( d_{i} \) for speaker (\(s\)) and accent (\(a\)). The reparametrized speaker \( z_{s} \) and grouped accent \( z^{g}_{a} \) representations pass through the VQ layer, acting as a bottleneck \cite{van2017neural}, filtering out irrelevant information. This integration improves accent conversion and preserves key information by effectively disentangling speaker and accent attributes. The VQ block incorporates an information bottleneck, ensuring effective utilization of codebooks. We define a latent embedding space $e^{i} \in \mathcal{R}^{d_{i}} \times D$, where $d_{i}$ represents the size of the discrete latent space, $i \in \{s,a\}$ denotes speaker and accent, and $D$ corresponds to the dimensionality of each latent embedding vector $e^{i}$. It is important to note that within this space, $d_{i}$ embedding vectors $e_{j}^{i} \in \mathcal{R}^{D}$ exist, where $j$ ranges from 1 to $d_{i}$. To ensure that the representation sequence effectively commits to an embedding and to prevent its output from growing, we incorporate a commitment loss, following prior research \cite{van2017neural}, for each VQ module. This loss helps in stabilizing the training process:
\begin{equation}
    \mathcal{L}_{c} = || z_{e^{i}}(x) - sg[e^{i}]||^{2}_{2}
\end{equation}

where $z_{e^{i}}(x)$ is the output of the $i^{th}$ vector quantization block ( $i \in \{s,a\}$), and $sg$ stands for the stop gradient operator. Finally, by adding the KL loss multiplied by coefficient $\beta$, the total loss for training is computed as:

\begin{equation}
    \mathcal{L}_{total} = \mathcal{L}_{recon} + \beta \mathcal{L}_{kl} + \mathcal{L}_{c}
    \label{eq:total}
\end{equation}

\section{Experimental Setup and Results}
\label{sec:experiment}
\subsection{Datasets and Baselines}
We use two datasets for training: the train-clean-100 subset of LibriTTS \cite{zen2019libritts} (LTS), and the L2-ARCTIC dataset \cite{zhao2018l2}. LTS includes $247$ English speakers, whereas the L2-ARCTIC dataset comprises $24$ L2 (second-language) speakers representing $6$ accents, with each accent having $4$ speakers (two females and two males). The evaluation is conducted on the L2-ARCTIC validation set.

We train the baselines and the proposed model using two strategies. First, we train the TTS system from scratch with accented data. Second, a two-step process where the TTS backbone is initially trained on an English-only corpus, yielding a pre-trained multispeaker backbone model which uses GE2E speaker embeddings~\cite{wan2018generalized}, and then fine-tuned with accented data. In this case, if the model uses GST or ML-VAE modules, they replace the GE2E speaker embeddings from pre-training. Details on training parameters and procedure can be found in \ref{app:trainpara}. We then evaluate and compare \model{}'s performance against various TTS architectures with both autoregressive and non-autoregressive frameworks. We define the baselines and different variants of the proposed model as follows:

\par{\textbf{Baselines: }\textbf{\baselineA{}} represents the TTS architecture proposed in \cite{melechovsky2023learning}. It consists of Tacotron2 with \encoder{} and is trained with L2-ARCTIC.
\textbf{\baselineD{}} is our pre-trained multispeaker FastSpeech2 backbone model, pre-trained on LTS, fine-tuned on L2-ARCTIC.}
\textbf{\baselineE{}} is a pre-trained multispeaker FastSpeech2 model with a GST to model speakers/accents.
\textbf{\baselineF{}} is a pre-trained multispeaker FastSpeech2 model with a GST to model accents and GE2E embeddings to model speakers.

\par{\textbf{\model{} versions: }\textbf{\baselineB{}} represents the proposed architecture, however, the entire model is trained from scratch on L2-ARCTIC.
\textbf{\baselineC{}} denotes the \model{} architecture that uses pre-trained multispeaker TTS as a backbone with \encoder{} but without Vector Quantization modules in \encoder{}.
\textbf{\model{}} leverages a pre-trained multi-speaker TTS system as its foundational architecture, enhanced by the \encoder{} and a Vector Quantization module, as illustrated in Figure \ref{fig:startts}.
To thoroughly assess the efficacy of our proposed approach, we also explored various versions of \model{} with differing codebook sizes $d_{i}$ of $512$, $128$, and $64$. Based on the objective results and human evaluation, we chose the size of $512$ for further experiments. More details in \ref{app:codebook}.}

\label{sec:backbone}
\subsection{Objective \& Subjective Evaluation}
\label{sec:obj_eval}

We evaluate timbre and prosody similarity between synthesized and reference audio using Cosine Similarity (CS) \cite{dehak2010front} and F0 Frame Error (FFE) \cite{talkin1995robust}, calculating average CS between embeddings from synthesized and ground truth data for speaker similarity, while FFE captures fundamental frequency information by combining voicing decision error and F0 error metrics. To assess perceptual dissimilarities, we use Mel Cepstral Distortion (MCD) \cite{mcd1}, which measures the divergence between the MFCCs of synthesized and original speech, and compute the WER \cite{wer} to measure speech intelligibility using enterprise-grade, pre-trained Silero speech-to-text.

To assess speech quality and accent-speaker disentanglement, we conducted subjective listening tests with two groups: AR and NAR. In the AR group, we compared different variants of \model{} with ground truth and speech generated from an autoregressive model (\baselineA{}). In the NAR group, we compared different variants of \model{} with ground truth and speech generated from non-autoregressive models (\baselineE{}, \baselineF{}). In each group, we evaluated naturalness through the Mean Opinion Score (MOS). Additionally, we aimed to evaluate the accent-speaker disentanglement by asking listeners to rate accent-converted samples on both speaker and accent similarity metrics. For this purpose, we performed Best-Worst Scaling (BWS) tests \cite{louviere2015best} in each group. The samples in the AR group are evaluated by $11$ listeners, whereas $12$ listeners evaluated the samples in the NAR group\footnote{The evaluators are NLP and speech processing researchers, who are familiar with subjective evaluation.}.

\begin{table}[h]
    \centering
    \caption{Objective \& Subjective evaluation. GT represents the ground truth.}
    \resizebox{0.8\textwidth}{!}{
    \begin{tabular}{@{}lcccc|cc|cc|@{}}
    \toprule
    \rowcolor[gray]{.9}\multicolumn{5}{c|}{\textbf{Objective Evaluation}} & \multicolumn{4}{c}{\textbf{Subjective Evaluation}} \\
    \midrule
    Metric & MCD $\downarrow$  & CS $\uparrow$ & FFE $\downarrow$ & WER  $\downarrow$ & MOS & 95\% CI & MOS & 95\% CI \\
    \midrule
    GT  & - & - & - & 0.137 & 4.510 & 0.154 & 4.498 & 0.219 \\ 
    \midrule
    \rowcolor[gray]{.9}\multicolumn{5}{c|}{\textbf{Autoregressive}} & \multicolumn{2}{c|}{\textbf{\model{} vs AR}} & &\\ 
    \midrule
    \baselineA{}   & 6.952 & 0.785 & 0.490 & 0.216 & 2.556 & 0.213 & - & - \\ 
    \midrule
    \rowcolor[gray]{.9}\multicolumn{5}{c|}{\textbf{Non-Autoregressive}} & & & \multicolumn{2}{c}{\textbf{\model{} vs NAR}}\\ 
    \midrule
    \baselineD{}  & 6.952 & \underline{0.850} & 0.426 & 0.174 & - & - & 3.335 & 0.297 \\ 
    \baselineE{}   & 6.731 & 0.841 & 0.438 & 0.184 & - & - & 3.286 & 0.246 \\
    \baselineF{}   & \textbf{6.571} & 0.828 & 0.449 & 0.187 & - & - & 3.615 & 0.185 \\
    \midrule
    \baselineC{}   & 6.731 & 0.841 & \underline{0.426} & 0.186 & 3.389 & 0.189 & - & - \\ 
    \model{}$_{512}$ & 6.934 & 0.842 & 0.431 & \textbf{0.164} & 3.150 & 0.162 & 3.346 & 0.133 \\
    \baselineB{}   & \underline{6.669} & \textbf{0.859} & \textbf{0.409} & \underline{0.169} & \textbf{3.484} & 0.214 & - & - \\ 
    \bottomrule
    \end{tabular}
    }
    \label{tab:obj_eval}
    \vspace{-3mm}
\end{table}


\subsection{Results and Discussion}
\textbf{Objective Evaluation:} Table~\ref{tab:obj_eval} presents the objective evaluation results for various baselines and \model{}. In both autoregressive (AR) and non-autoregressive (NAR) baselines, \model{} demonstrates improvement across all metrics. Furthermore, \baselineB{} achieves the best overall scores in various objective metrics, e.g., the highest speaker cosine similarity score of $0.859$, demonstrating the effectiveness of the proposed architecture in multispeaker speech synthesis.



\begin{figure*}[htp]
  \centering
  \subfigure[Accent similarity: \model{} vs AR]{\includegraphics[scale=0.18]{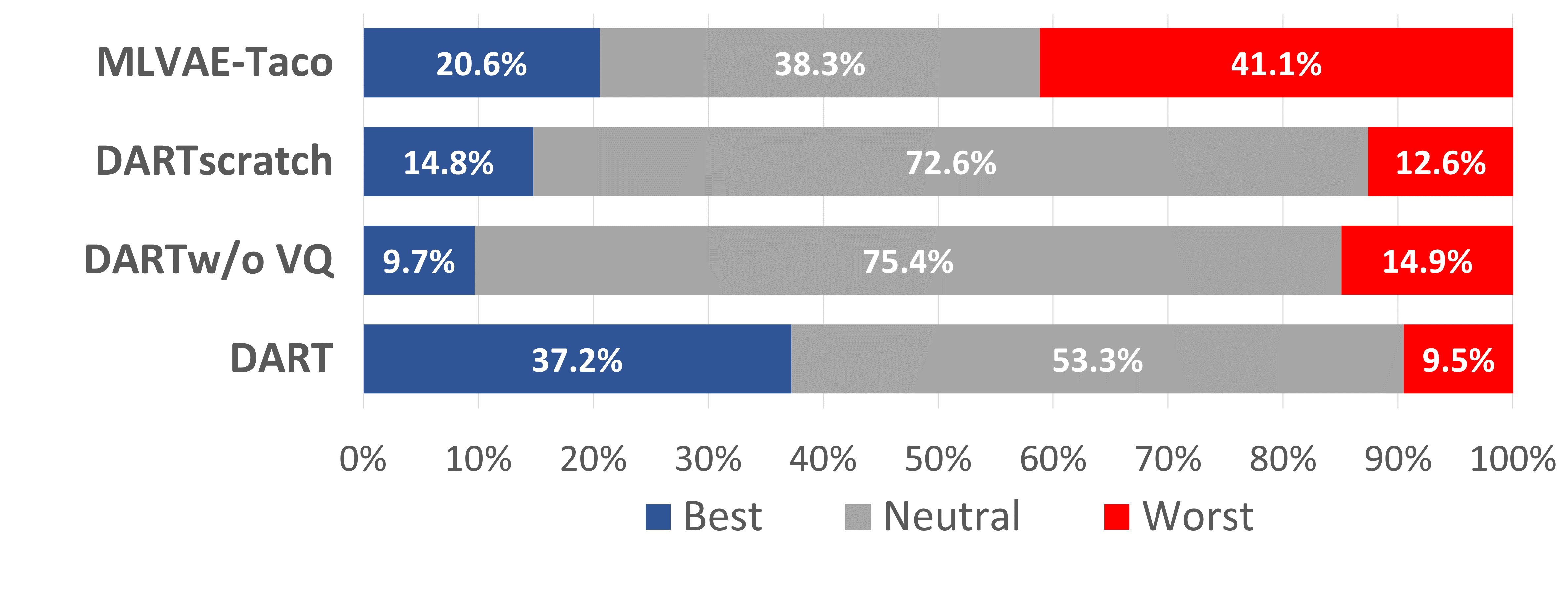}\label{fig:acc_sim}}\quad
  \subfigure[Speaker similarity: \model{} vs AR.]{\includegraphics[scale=0.18]{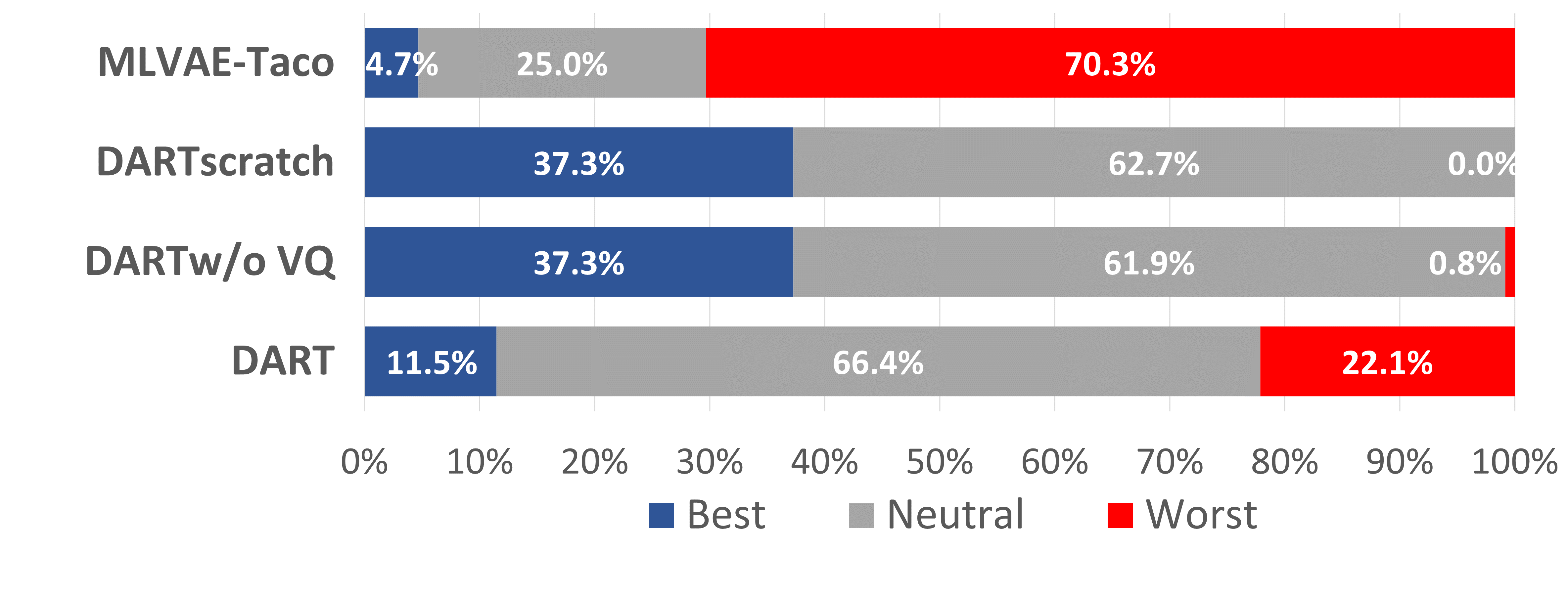}\label{fig:spk_sim}}
  \subfigure[Accent similarity: \model{} vs NAR]{\includegraphics[scale=0.18]
  {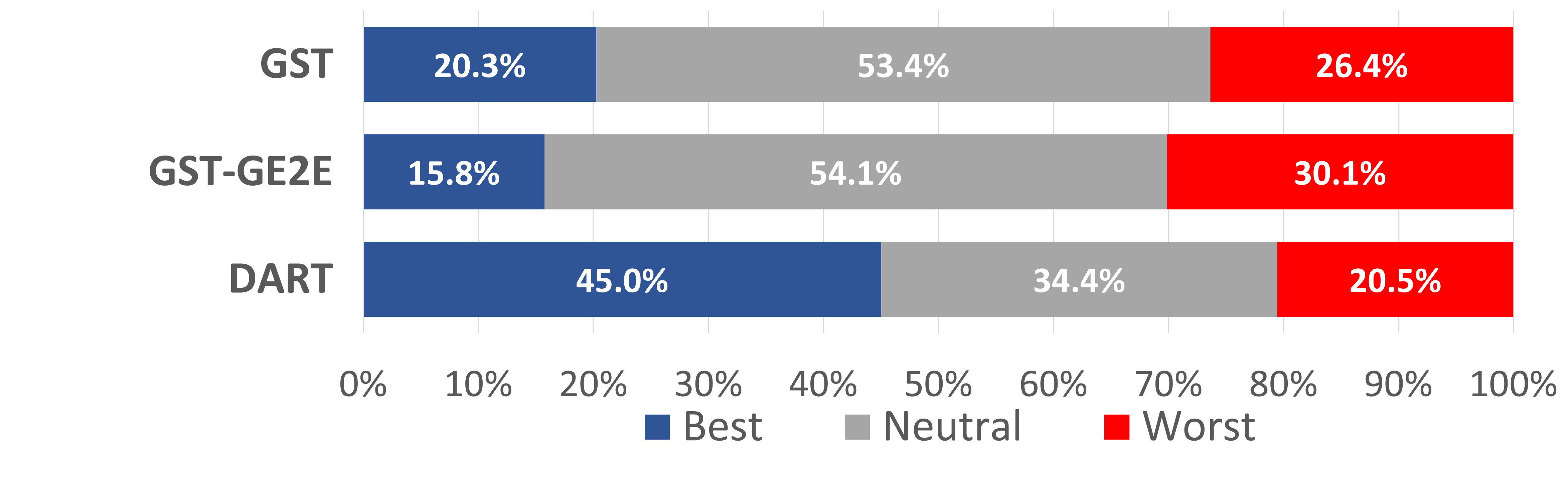}\label{fig:acc_sim2}}\quad
  \subfigure[Speaker similarity: \model{} vs NAR.]{\includegraphics[scale=0.18]
  {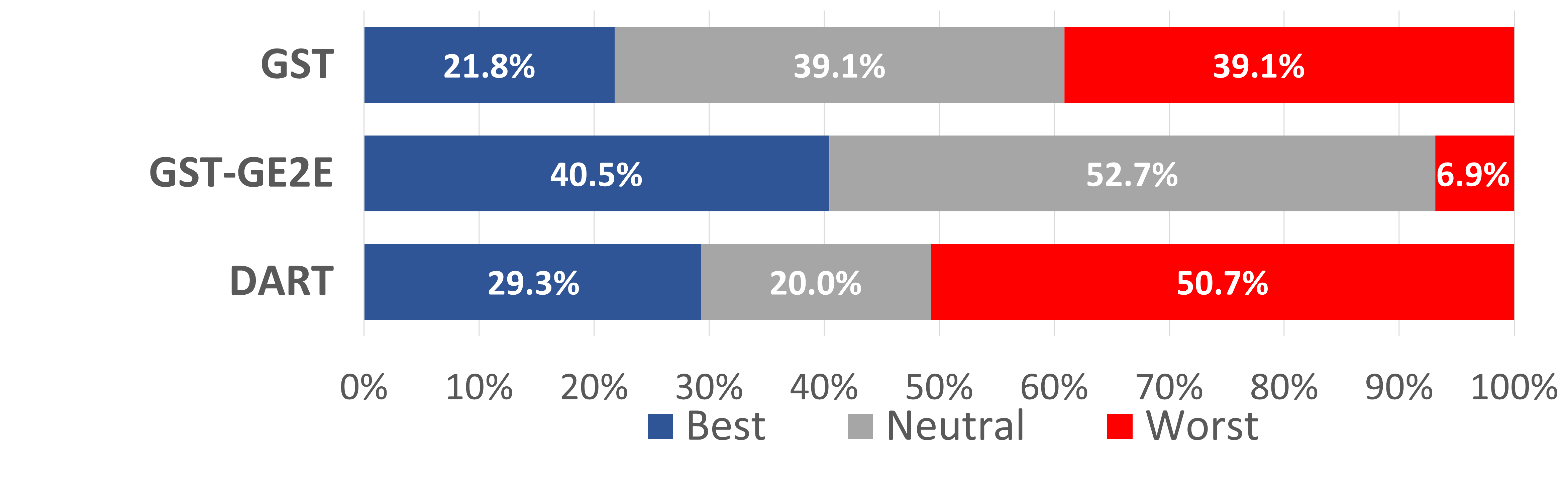}\label{fig:spk_sim2}}
  \vspace{-2.5mm}
  \caption{Subjective evaluation: Best-Worst-Scaling (BWS).}
   \label{fig:BWSresult2}
    \vspace{-5.5mm}
\end{figure*}



\textbf{Subjective Evaluation:} We perform comprehensive subjective evaluation as discussed in Section \ref{sec:obj_eval} to assess the effectiveness of \model{} for accent conversion. In the subjective evaluation results for AR group (Table~\ref{tab:obj_eval}, Fig.~\ref{fig:BWSresult2}), we can observe that all variants of \model{} outperform \baselineA{} in naturalness and in speaker and accent similarity. Interestingly, among the different variants, \model{}${_{512}}$ exhibits a slightly lower MOS score of $3.150$ compared to both \baselineB{} and \baselineC{}. Further analysis of the results reveals that \baselineB{} and \baselineC{} perform poorly in accent similarity, they achieve higher speaker similarity than \model{}${_{512}}$ (Figure \ref{fig:acc_sim} \& \ref{fig:spk_sim}). This observation supports our hypothesis that leveraging a pretrained TTS backbone and VQ aids in the disentanglement with slight dip in overall naturalness.

Similarly, the MOS (Table~\ref{tab:obj_eval}) and BWS (Figures \ref{fig:acc_sim2} \& \ref{fig:spk_sim2}) results for the NAR group further validate our claim that \model{}$_{512}$ effectively disentangles accent and speaker representation. This is evident in Figure \ref{fig:acc_sim2}, where speech samples generated using \model{}$_{512}$ are preferred $45.0\%$ of the time for accent similarity over \baselineE{} and \baselineF{}. It is important to note that the higher speaker similarity score of \baselineF{} is due to its use of state-of-the-art speaker embedding~\cite{wan2018generalized} for generating speech samples. Additionally, \baselineE{} fails to effectively capture both accent and speaker characteristics, as highlighted by the results in Figures~\ref{fig:acc_sim2} \& \ref{fig:spk_sim2}. Furthermore, we want to highlight that the lower performance in the MOS score for \model{}$_{512}$ can be attributed to several factors related to the trade-offs between accent and speaker similarity. Achieving a perfect balance between maintaining speaker identity and accurately converting the accent may result in slightly compromised overall naturalness, as reflected in the MOS.

 \textbf{Importance of pre-training}: We observe that \baselineB{} and \model{}$_{512}$ share the same architecture but differ in training strategy. \baselineB{} is trained from scratch (Backbone \& \encoder{}), while in \model{}$_{512}$, the backbone TTS was first pre-trained on a multi-speaker English corpus, followed by training the \encoder{} with accent data alongside the TTS backbone. Although \baselineB{} demonstrates better performance in objective metrics, \model{}$_{512}$ performs significantly better in accent conversion.
 We attribute this to \model{}$_{512}$'s prior knowledge of many different voices, gained through pre-training, as accent-converted voices represent new, unseen voices.
 Thus, there is a trade-off in designing speech synthesis systems that specifically account for accents. Pre-training can enhance accent conversion at the cost of a slight reduction in perceived identity. In future work, we aim to bridge this gap and simultaneously improve both accent conversion and perceived identity.

\section{Conclusion}
\label{sec:conclusion}
Our proposed approach significantly enhances the capabilities of multispeaker TTS models by effectively disentangling speaker and accent representations, resulting in more flexible and personalized speech synthesis. This has broad applications in entertainment; personalization of virtual assistants, narrators; and more. By utilizing \encoder{} and VQ, our proposed method achieves superior accent conversion. In future work, we will focus on further advancing the disentanglement between speaker and accent in multispeaker TTS. This includes addressing the trade-off between disentanglement and naturalness, expanding datasets and exploring real-time zero-shot adaptation techniques.

\begin{ack}
This project has received funding from SUTD Kickstarter Initiative no. SKI 2021\_04\_06.\\
The work by Berrak Sisman was funded by NSF CAREER award IIS-2338979.
\end{ack}

\bibliographystyle{unsrt} 
\bibliography{references}    








\newpage
\appendix

\section{Appendix / supplemental material}
\subsection{Related works in accented speech synthesis}
\label{app:relworks}

There are a few related works in accented speech synthesis. In Voice Conversion (VC) \cite{sisman2020overview,10030068}, the sub-field of Foreign Accent Conversion (FAC) focuses on transforming L2 speaker's voice to an L1 native accent. It is oftentimes used for applications like Computer-Assisted Pronunciation Training (CAPT) \cite{felps2009foreign,rogerson2021computer}, which aims to help L2 speakers improve their pronunciation towards a more native-sounding one. Early FAC methods used spectral or cepstral features from native and non-native speakers \cite{aryal2013foreign,felps2010developing,felps2009foreign,huckvale2007spoken}, while others incorporated phonetic posteriorgrams (PPGs) to capture phonetic variations \cite{sun2016phonetic,10030068,zhao2019foreign}. Recent advancements leverage deep learning; Wang et al. used adversarial learning to disentangle accent and speaker identity \cite{wang2021accent}, and Accentron employed ResNet-34 classifiers for accent and speaker recognition \cite{DING2022101302}.
In TTS, some approaches treat accent as a style component, as seen in GMVAE-Tacotron \cite{hsu2018hierarchical} and GST-Tacotron \cite{wang2018style}. Liu \textit{et al.} improved L2 accent intensity using a variance adaptor \cite{liu2024controllable}. Melechovsky \textit{et al.} proposed disentangling accent and speaker with ML-VAE and Tacotron2 \cite{melechovsky2022accented,melechovsky2023learning}, achieving results comparable to GMVAE \cite{tan2019hierarchical,hsu2018hierarchical}, while other works \cite{10173587} enhanced encoder-decoder frameworks with accent ID conditioning for varied phoneme representations.

\subsection{Training parameters}
\label{app:trainpara}
Here, we brielfy describe the training parameters used in our experiments. The FastSpeech2~\cite{ren2020fastspeech} TTS backbone follows the original architecture with a hidden state dimension of $256$.
During training, the speaker embeddings are added to the text representation in the variance adapter.
Speaker embeddings are computed using a speaker verification model trained with the GE2E loss~\cite{wan2018generalized}, incorporating LibriSpeech (train-other-500), Voxceleb1, and Voxceleb2 datasets~\cite{panayotov2015librispeech}, Voxceleb1, and Voxceleb2 \cite{nagrani2017voxceleb}.
To improve model convergence for unsupervised duration modeling, variance adapter training begins at 50K steps. All speech samples are downsampled to $16$ kHz. The models are trained using the Adam optimizer with $4$K warmup steps, followed by annealing at $300$K, $400$K, and $500$K steps, and a total training duration of $600$K steps.

The \encoder{} along with the TTS backbone is fine-tuned using L2-ARCTIC \cite{zhao2018l2}, with the text encoder frozen while updating the weights of the variance adapter, mel-decoder, and \encoder{} module. KL loss coefficient $\beta$ in Eq \ref{eq:total} is set to $10^{-4}$. The fine-tuned models, including \model{} and \baselineC{}, undergo training for $100K$ steps to achieve convergence. Meanwhile, the baseline models \baselineA{} and \baselineB{}, constructed from scratch, are trained for $200K$ steps to reach convergence.
\begin{table}[h]
    \centering
    \caption{Comparative analysis of codebook sizes.}
    \begin{tabular}{@{}lcccc@{}}
    \toprule
    Metric & MCD $\downarrow$  & CS $\uparrow$ & FFE $\downarrow$ & WER $\downarrow$ \\
    \midrule
        \model{}$_{64}$ & 6.972 &0.842  & \textbf{0.425} &\textbf{0.161}\\
    \model{}$_{128}$ & 7.019 &0.842 & 0.427 & 0.168\\
        \model{}$_{512}$ & \textbf{6.934} & \textbf{0.842} & 0.431 & 0.164\\
 \bottomrule
    \end{tabular}
    \label{tab:obj_eval_codebook}
\end{table}

\subsection{Effect of VQ codebook size}
\label{app:codebook}
Here, we present the objective results for our VQ codebook size experiment, as seen in Table \ref{tab:obj_eval_codebook}. We observe that while \model{}$_{512}$ achieves the best performance in MCD and CS, \model{}$_{64}$ demonstrates superior performance in FFE and WER. This indicates that the choice of codebook size involves balancing various objective metrics, with smaller sizes favoring some metrics and larger sizes favoring others.
Since the differences in FFE and WER were negligible and we aimed to select a model with low distortion and high speaker similarity, we used \model{}$_{512}$ for subjective evaluation.

\subsection{Plottings of accent and speaker embeddings}
\label{app:embeddings}

\begin{figure*}[htp]
  \centering
  \subfigure[Accent embeddings without grouping $z_{a}$]{\includegraphics[scale=0.2]{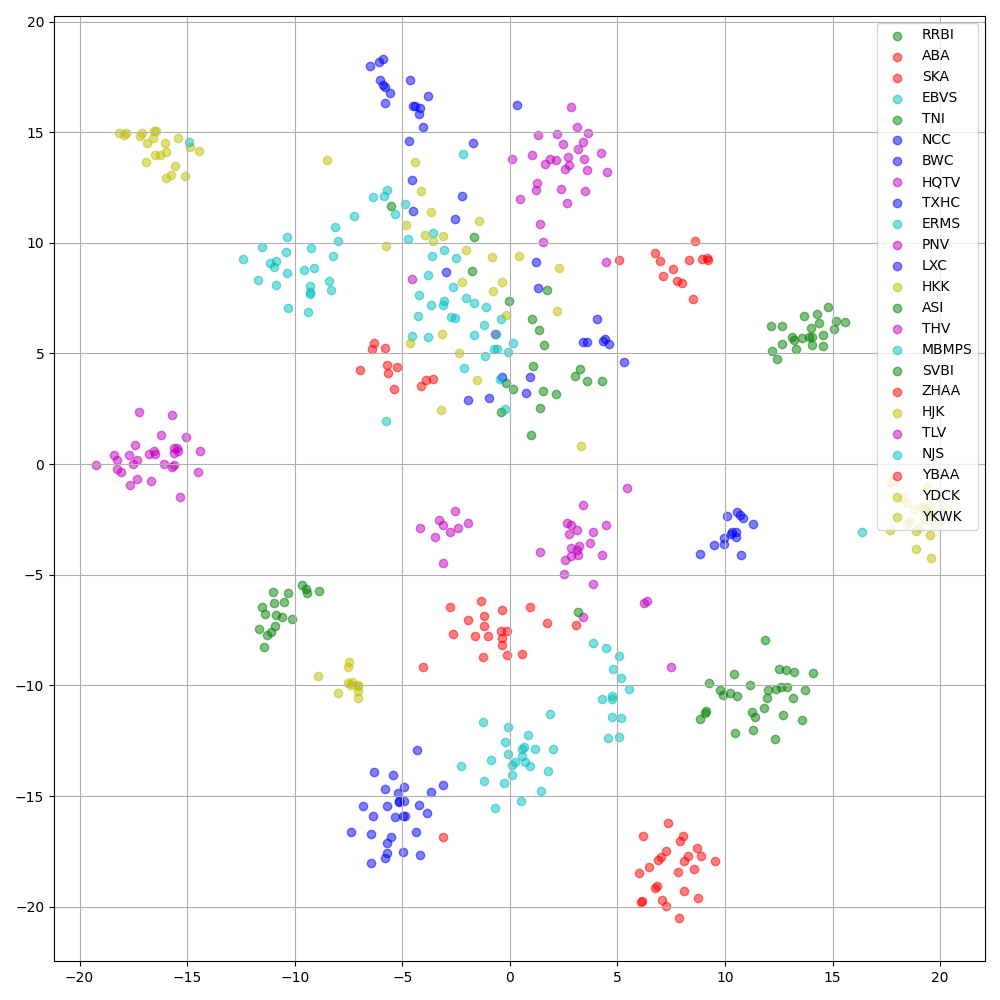}}\quad
  \subfigure[Accent embeddings with grouping $z_{a}^{g}$]{\includegraphics[scale=0.2]{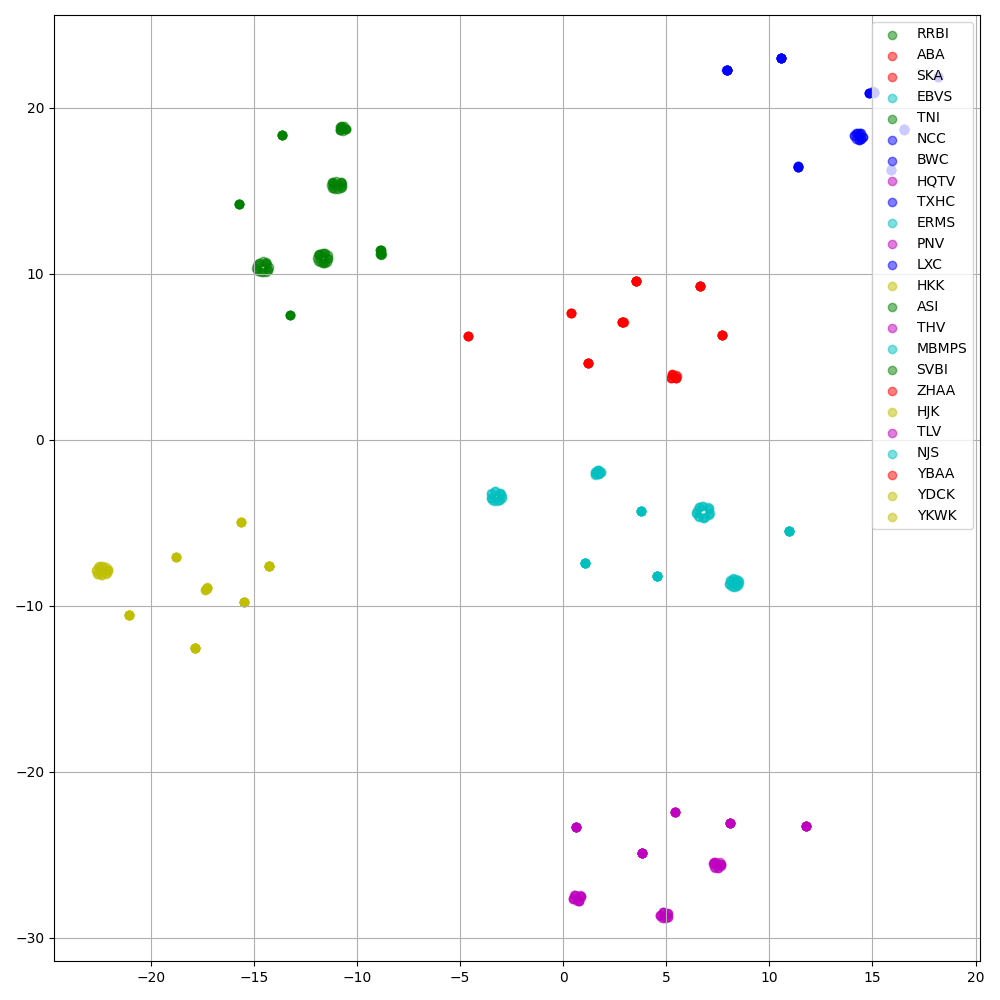}}
  \subfigure[Speaker embeddings $z_{s}$]{\includegraphics[scale=0.2]{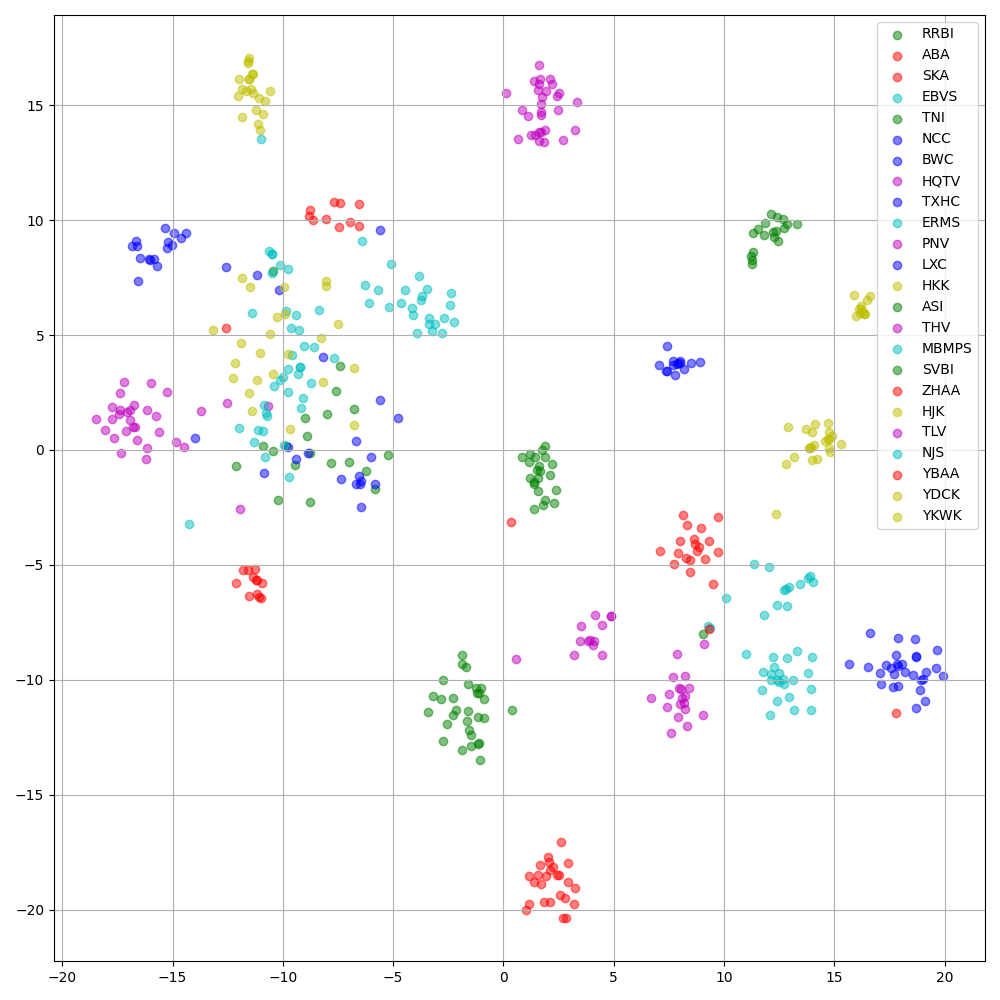}}
  \caption{The t-SNE plot demonstrating effective clustering and disentanglement of accent and speaker embeddings.}
  \label{fig:embeddings}
\end{figure*}

The t-SNE plots for accent and speaker embeddings from \model{}$_{512}$, shown in Figures \ref{fig:embeddings} (a) and (b), further illustrate the effectiveness of the \encoder{} and VQ modules in capturing and clustering accent representations across various speakers. Furthermore, Figure \ref{fig:embeddings} (c) visualizes the model's capability to disentangle accent (depicted by different colors) and speaker attributes, resulting in enhanced accent conversion and robust accent representation.

\end{document}